\documentclass{article}
\usepackage{spconf,amsmath,graphicx,hyperref}
\usepackage{cite}
\usepackage{amsmath,amssymb,amsfonts}
\usepackage{algorithmic}
\usepackage{graphicx}
\usepackage{textcomp}
\usepackage{xcolor}
\usepackage{multirow}
\usepackage{comment}
\usepackage{booktabs}
\usepackage{tikz}
\usepackage{pgfplots}
\usepackage{hyperref} 
\usepackage{cleveref}
\usepgfplotslibrary{groupplots}
\pgfplotsset{compat=1.18}  
\usepackage{siunitx}

\usepackage{mathtools}
\usepackage{enumerate}
\usepackage{enumitem}
\usepackage{siunitx}
\usepackage{tabularx}
\usepackage{array}
\usepackage{makecell}

\usepackage[utf8]{inputenc}
\usepackage[T1]{fontenc}

\def\BibTeX{{\rm B\kern-.05em{\sc i\kern-.025em b}\kern-.08em
    T\kern-.1667em\lower.7ex\hbox{E}\kern-.125emX}}

\title{AlignAudio: A Unified Dual-Alignment Framework for Noise-Robust Speech-Conditioned Generation}

\name{\textbf{Yixuan Li}$^{1, 2}$ \quad \textbf{Xuenan Xu}$^{3}$ \quad \textbf{Zeyu Xie}$^{4}$ \quad \textbf{Mengyue Wu}$^{1}$\sthanks{
    Corresponding author.
  }}
\address{
    $^1$ X-LANCE Lab, Shanghai Jiao Tong University, China \qquad $^2$ Shanghai AI Lab, China \\
    $^3$ KlingAI Research, China \qquad $^4$ Peking University, Shenzhen, China
}

\begin{document}
\ninept
\setlength{\textfloatsep}{10pt plus 2pt minus 2pt}
\setlength{\dbltextfloatsep}{10pt plus 2pt minus 2pt}
\setlength{\intextsep}{8pt plus 2pt minus 2pt}

\setlength{\abovedisplayskip}{4pt plus 1pt minus 1pt}
\setlength{\belowdisplayskip}{4pt plus 1pt minus 1pt}
\setlength{\abovedisplayshortskip}{2pt plus 1pt minus 1pt}
\setlength{\belowdisplayshortskip}{2pt plus 1pt minus 1pt}

\maketitle

\begin{abstract}
Speech-conditioned generation maps speech signals to target audio outputs. Prior end-to-end speech-conditioned generative models achieve high fidelity in clean acoustic conditions, but their performance degrades substantially under environmental noise. While prior noise-robust approaches have achieved significant success in feature-level purification, maintaining consistent generation under clean and noisy conditioning remains a key challenge. To address this, we propose AlignAudio, a noise-robust generative framework based on dual alignment: (i) aligning the representations of clean and noisy speech to preserve semantic cues, and (ii) enforcing generation-level consistency within a flow matching paradigm to improve conditioning fidelity. Experiments on speech-to-audio (STA) and voice conversion (VC) tasks demonstrate that AlignAudio maintains performance comparable to baselines on clean speech while significantly outperforming them across various noisy conditions. Furthermore, the proposed dual-alignment strategy is applicable to both single- and multi-stream speech conditioning and can be integrated into different flow matching-based generation frameworks and post-training settings. Audio demonstrations and qualitative comparisons: \href{https://alignaudio.github.io/AlignAudio}{\textcolor{cyan}{\textit{https://alignaudio.github.io/AlignAudio}}}.
\end{abstract}

\vspace{-1em}

\section{Introduction}
\label{sec:intro}

Recent years have witnessed remarkable progress in speech and audio generation, driven by advances in diffusion models \cite{ho2020denoising}, flow matching \cite{lipman2023flow}, and language modeling approaches~\cite{borsos2023audiolm}. These generative techniques have enabled a new paradigm of speech-conditioned generation, where spoken input is directly mapped to diverse audio outputs, encompassing tasks such as speech-to-audio (STA) generation that synthesizes environmental sounds from spoken descriptions~\cite{xie2025starspeechtoaudiogenerationrepresentation}, and voice conversion (VC) that transforms speaker identity while preserving linguistic content~\cite{liu2024zero, li2023freevc}.

However, most studies in speech-conditioned audio generation assume ideal clean-speech conditions, where both training and evaluation are conducted on carefully curated datasets with minimal environmental interference.
In real-world scenarios, speech signals are inevitably corrupted by background noise, reverberation, and competing acoustic events. Previous studies on voice conversion have shown that noisy source or reference speech significantly degrades speaker similarity and speech naturalness~\cite{he2025noro,jiang2025refvc}.
Our preliminary experiments on speech-to-audio generation also demonstrate that environmental noise significantly degrades generation performance, increasing Fréchet Audio Distance (FAD) by approximately 33\% and reducing Contrastive Language-Audio Pretraining (CLAP) score  by around 18\%, indicating degraded audio quality and weaker semantic alignment.
As a result, speech-conditioned generation under realistic acoustic interference remains insufficiently studied, motivating the need for noise-robust speech-conditioned generation.

\begin{figure}[t]
  \centering
  \includegraphics[width=\linewidth]{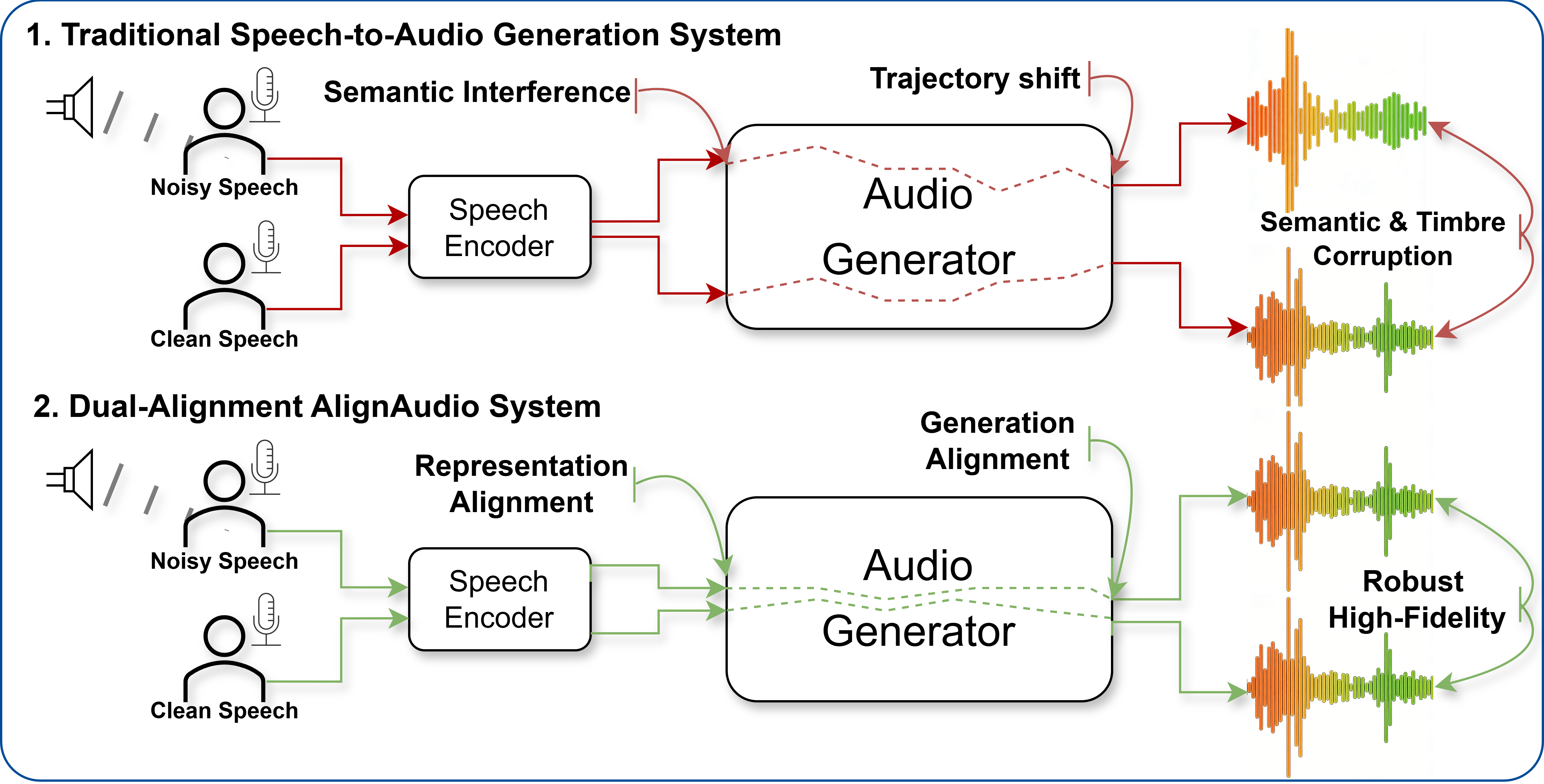}
  \vspace{-2em}
  \caption{AlignAudio generates audio with better noise robustness, preserved acoustic events, and higher semantic fidelity compared to traditional speech-conditioned generation systems that suffer from semantic interference and trajectory shifts under noise.}
  \label{fig:intro}
\end{figure}

\begin{figure*}[t]
  \centering
  \includegraphics[width=\linewidth]{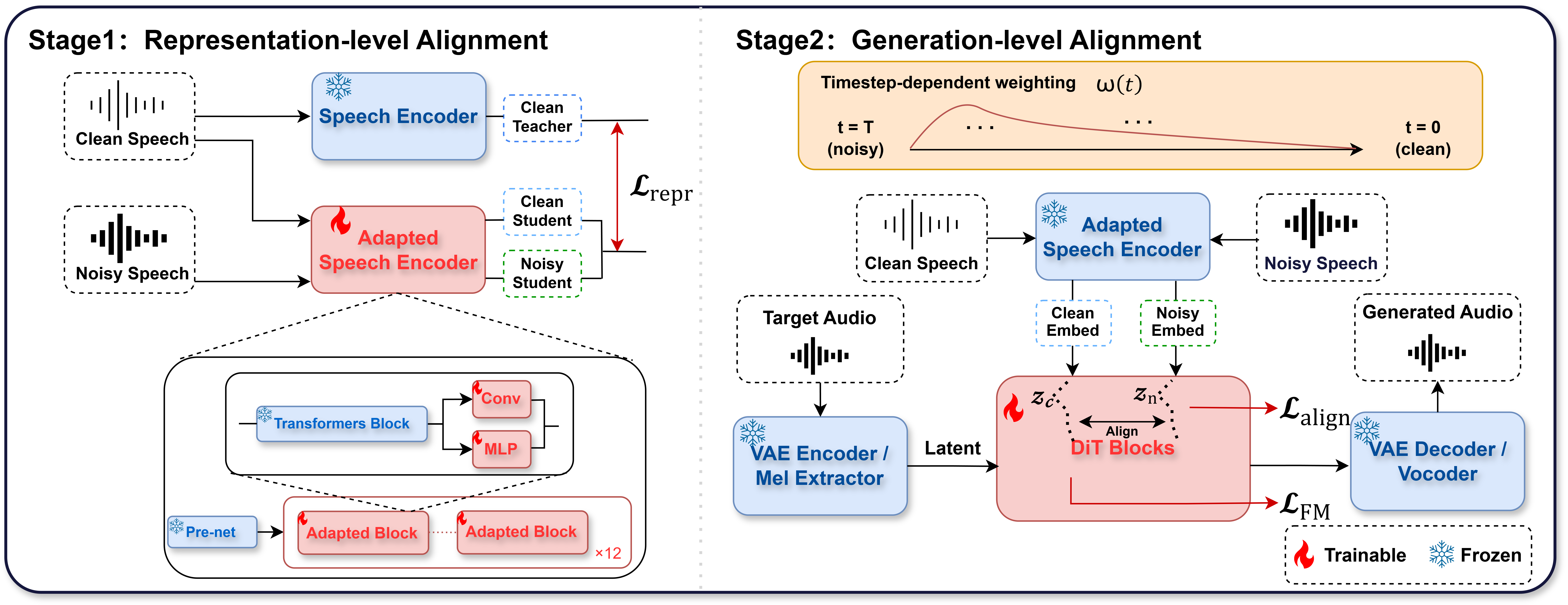}
  \vspace*{-2em}
  \caption{Overview of the AlignAudio architecture. Representation-level alignment (Stage 1) uses lightweight adapters in a pre-trained speech encoder to distill noisy speech embeddings toward clean ones via $L_2$ loss, while generation-level alignment (Stage 2) regularizes the noisy-conditioned velocity field in DiT blocks toward the clean one using directional similarity.}
  \label{fig:AlignAudio}
\end{figure*}

Several lines of research have addressed noise robustness in speech-conditioned generation, primarily through front-end speech enhancement (SE)~\cite{lu2022study,richter2023speech} or noise-invariant representation learning inspired by robust automatic speech recognition (ASR)~\cite{park2019specaugment,baevski2020wav2vec}.
These methods improve the robustness of conditioning signals and have proven effective for tasks such as noise-robust voice conversion~\cite{zhang2021denoispeech,hsu2019disentangling}.
Noro~\cite{he2025noro} improves robustness to noisy speaker references through dual-branch reference encoding and a noise-agnostic contrastive speaker loss.
REF-VC~\cite{jiang2025refvc} addresses noisy source speech by combining ASR and self-supervised features with random erasing to balance noise robustness and expressiveness, while implicit input--output alignment reduces reconstruction of irrelevant source information.

Despite these advances, these methods primarily address corruption in a specific conditioning stream.
As speech-conditioned generation increasingly relies on multiple heterogeneous conditioning signals (e.g., the source content and target speaker reference in voice conversion), perturbations from different inputs can jointly propagate through the shared generation trajectory, making representation-level robustness alone insufficient.
Moreover, neither method explicitly enforces consistency between clean- and noisy-conditioned generation dynamics.
Even with noise-robust conditioning features, the conditional generation dynamics (e.g., the velocity field in flow matching models) can remain sensitive to residual perturbations. These perturbations can accumulate throughout the generation process, resulting in degraded acoustic quality, reduced conditioning fidelity, and inaccurate semantic generation.

Instead of solely enhancing noisy speech inputs or learning noise-robust conditioning representations, we propose AlignAudio, a dual-alignment framework that explicitly regularizes both conditioning representations and conditional generation dynamics.
Representation-level alignment reduces the distribution gap between clean and noisy conditioning signals through lightweight adapters in the speech encoder or multi-input fusion module, providing robust features for both single- and multi-stream speech conditioning.
Building upon these robust representations, generation-level alignment further constrains the conditional trajectory to follow consistent dynamics under clean and noisy settings.
By jointly improving robustness at both levels, AlignAudio mitigates trajectory instability while naturally generalizing to diverse conditioning configurations (Fig.~\ref{fig:intro}).
We validate its effectiveness on two tasks: STA and VC.

Overall, our contributions are summarized as follows:
\begin{itemize}[
  leftmargin=1em,
  topsep=2pt,
  partopsep=0pt,
  parsep=0pt,
  itemsep=0.2pt
]
\item We propose AlignAudio, a dual-alignment framework for noise-robust speech-conditioned generation. It accommodates both single-stream (STA) and multi-stream (VC) speech conditioning, mitigating noise at both the representation and generation levels.

\item We demonstrate the flexibility of AlignAudio across different flow-based generation paradigms, showing that it extends from standard Flow Matching to efficient MeanFlow training and further integrates with Flow-GRPO post-training for improved content preservation.

\item Extensive experiments show that AlignAudio achieves substantial improvements over strong baselines on both STA and VC in conditioning fidelity under noisy conditions. It exhibits strong noise robustness and robust generalization to challenging out-of-domain (OOD) conditions, while maintaining competitive performance on clean speech.

\end{itemize}

\vspace{-1em}

\section{Noise-Robust Generation via Dual-Alignment}
\label{sec:method}

As illustrated in
Fig.~\ref{fig:AlignAudio}, AlignAudio consists of two complementary modules: represen\-tation-level alignment and generation-level alignment, which improve robustness by aligning conditioning representations and generation trajectories, respectively.
\vspace{-1em}
\subsection{Preliminaries}

Following recent advancements in efficient audio generation~\cite{xu2025uniflow}, we adopt flow matching as our base generative framework. Standard flow matching trains a model to predict the \textit{instantaneous velocity} $v_\theta(x_t, t)$ to transport a simple prior $x_0 \sim p_0$ to the target data $x_1 \sim p_1$ via a linear path $x_t = (1-t)x_0 + tx_1$. The base objective is:
\begin{equation}
\label{eq:fm_base}
\mathcal{L}_{\text{FM}}(\theta)
= \mathbb{E}_{t, x_1, x_t}
\left[
\lVert v_\theta(x_t, t) - (x_1 - x_0) \rVert_2^2
\right].
\end{equation}

For an interval \([r,t]\), MeanFlow~\cite{geng2025meanflows} defines the average velocity as the displacement divided by the interval length, and derives the intrinsic identity
\begin{equation}
\label{eq:meanflow_identity}
\bar{\mathbf{v}}_{r,t}(\mathbf{z})
= \mathbf{v}_t(\mathbf{z})
- (t-r)\frac{d}{dt}\bar{\mathbf{v}}_{r,t}(\mathbf{z})
\end{equation}
which yields a principled training target through a Jacobian-Vector Product (JVP) term.
By directly predicting this interval-averaged field, MeanFlow enables single- or few-step sampling with strong generation quality. The corresponding predictor is denoted by $u_\theta(x_t,r,t)$.
\vspace{-1em}
\subsection{Representation-level Alignment}

Drawing on parameter-efficient adaptation~\cite{houlsby2019parameter,pfeiffer2020adapterhub,pfeiffer2020mad}, we introduce lightweight adapters into the conditioning pathways rather than fine-tuning the entire network. These adapters are trained via knowledge distillation, using representations from clean speech as soft targets.

Formally, let $E(\cdot)$ denote the frozen, clean-trained conditioning network (the teacher), and $A(\cdot)$ the corresponding network augmented with trainable adapters (the student). To extract noise-invariant features, we minimize the $L_2$ distance between the adapter outputs on noisy ($x_{\mathrm{n}}$) and clean ($x_{\mathrm{c}}$) inputs, and the clean teacher embeddings $E(x_{\mathrm{c}})$:
\begin{equation}
\label{eq:repr}
\begin{aligned}
\mathcal{L}_{\mathrm{repr}}
={}&(1-\lambda)\,
\lVert E(x_{\mathrm{c}})-A(x_{\mathrm{n}})\rVert_2^2\\
&+\lambda\,
\lVert E(x_{\mathrm{c}})-A(x_{\mathrm{c}})\rVert_2^2.
\end{aligned}
\end{equation}
where $\lambda \in [0,1]$ controls the trade-off between noisy and clean adaptation.

Adapter placement (e.g., within the frontend encoder or multi-input fusion module) is tailored to the specific downstream architecture. The adapted conditioning network is frozen during generation-level alignment.
\vspace{-1em}
\subsection{Generation-level Alignment}

\textbf{Trajectory Alignment.} Let $f_\theta \in \{v_\theta(x_t,t),\allowbreak u_\theta(x_t,r,t)\}$ denote the predicted field depending on the chosen backbone. We observe that even with representation-level alignment, noisy conditioning vectors can still induce subtle deviations in the generated trajectories. A straightforward pointwise $L_2$ alignment between the noisy-conditioned output $f_{\theta}^{\text{n}}$ and clean-conditioned output $f_{\theta}^{\text{c}}$ would enforce strict magnitude agreement, which can overly constrain the trajectory dynamics and degrade fine-grained acoustic details.

To preserve trajectory scale while ensuring directional consistency, we instead adopt a cosine directional alignment. We treat the clean flow as a fixed reference (with stop-gradient, $\textrm{sg}(\cdot)$) to prevent noisy samples from corrupting the clean generation capabilities:
\begin{equation}
\label{eq:cos}
\mathcal{L}_{\mathrm{cos}}
=\mathbb{E}\!\left[
\omega(t)\left(
1-\frac{
\langle f_\theta^{\mathrm{n}},
\operatorname{sg}(f_\theta^{\mathrm{c}})\rangle
}{
\lVert f_\theta^{\mathrm{n}}\rVert_2\,
\lVert\operatorname{sg}(f_\theta^{\mathrm{c}})\rVert_2
}
\right)\right].
\end{equation}
The clean and noisy predictions share the same state and time samples, with their arguments omitted for brevity.

\textbf{Timestep-Dependent Weighting.} For standard flow matching, early timesteps ($t \approx 0$) require stronger regularization to stabilize the coarse trajectory, while late timesteps ($t \approx 1$) demand reduced intervention to avoid over-regularization. Therefore, we introduce a timestep-dependent weighting strategy $\omega(t)$ to adaptively attenuate the alignment strength:
$\omega(t) = \frac{1}{1 + \exp(-(b - t)k)}, \quad t\in [0, 1]$.
Here, $b$ and $k$ control the inflection point and slope of the weighting curve. When utilizing MeanFlow, the temporal weighting is disabled. 
($\omega(t) \equiv 1$).

The overall supervised generation loss integrates the generative base loss ($\mathcal{L}_{\text{FM}}$ or $\mathcal{L}_{\text{MeanFlow}}$), computed under clean conditioning, with the alignment regularization:
\begin{equation}
\label{eq:gen}
\mathcal{L}_{\mathrm{gen}}
=\mathcal{L}_{\mathrm{base}}^{\mathrm{c}}
+\alpha\mathcal{L}_{\mathrm{cos}},
\end{equation}
where $\alpha > 0$ scales the alignment loss. Here, $\mathcal{L}_{\mathrm{base}}^{\mathrm{c}}$ uses the clean condition $A(x_{\mathrm{c}})$; stop-gradient is applied to the clean reference only in $\mathcal{L}_{\mathrm{cos}}$.
\vspace{-1em}
\subsection{Task Instantiations}

To comprehensively validate the effectiveness of AlignAudio under different conditioning complexities, we consider two representative speech-conditioned generation tasks: STA and VC, corresponding to single- and multi-stream speech conditioning, respectively.

\textbf{STA.} The input $x$ denotes a single speech sequence. We instantiate the conditioning network using the pre-trained SpeechT5 encoder~\cite{ao2022speecht5}. To implement the representation-level alignment, we insert our lightweight dual-path adapters (comprising depthwise temporal convolutions and channel-wise bottleneck MLPs) directly after the final Transformer layers of SpeechT5. Following EzAudio~\cite{hai2024ezaudio}, we employ a variational autoencoder (VAE) to map the target audio waveform into a compact continuous latent space. The generative core is a Diffusion Transformer (DiT). Conditioned on the noise-invariant representations from the speech encoder, it predicts the velocity field via flow matching.

\textbf{VC.} Unlike STA, the input $x$ is a dual-stream composition consisting of a source utterance $x_{\text{src}}$  and a target prompt $x_{\text{tgt}}$ (providing the desired timbre). The core architecture adopts the framework of MeanVC~\cite{ma2025meanvc}, where source bottleneck features are fused with reference timbre information and, together with speaker embeddings, condition a MeanFlow-based DiT decoder to generate the converted mel-spectrogram. For VC, adapters are applied to the multi-input fusion module.

\textbf{VC post-training.} We further adapt Flow-GRPO~\cite{liu2025flowgrpo} after the two-stage training. We optimize a composite reward that jointly encourages content preservation, speaker similarity, and perceptual naturalness. We use the Mega-ASR~\cite{xie2026megaasr} reward as the content-consistency signal, combining Word Error Rate (WER)-aware feedback with dynamic granularity for hard samples. We complement it with Speaker Encoder Cosine Similarity (SECS) to measure speaker similarity, and UTMOSv2~\cite{baba2024utmosv2} as a proxy for utterance-level naturalness.
Following this design, the final reward is written as
\begin{equation*}
R = \lambda_{\text{asr}} R_{\text{Mega-ASR}}
+ \lambda_{\text{spk}} R_{\text{SECS}}
+ \lambda_{\text{mos}} R_{\text{UTMOSv2}}.
\end{equation*}

\section{Experiments and results}

\begin{table*}[t]
\centering
\setlength{\tabcolsep}{1.6pt}
\renewcommand{\arraystretch}{1.05}
\caption{Speech-to-audio (STA) and voice conversion (VC) results.
\textbf{Bold} and underlined entries indicate the best and second-best speech-conditioned results for each task, condition, and metric, excluding EzAudio with ground-truth (GT) text.
AlignAudio-VC is the full model with Flow-GRPO; * denotes its variant without GRPO.
Recorded uses a separate test set of AudioCaps captions read by participants in real-world scenarios.
WER and SECS are reported in percent; $\downarrow$/$\uparrow$ indicate lower/higher is better.
\textendash{} denotes unavailable results.}
\label{tab:main_results_combined}
\begin{tabular*}{\textwidth}{@{\extracolsep{\fill}}l*{11}{c}@{}}
\toprule
System & \multicolumn{3}{c}{Clean} & \multicolumn{3}{c}{Noisy} & \multicolumn{3}{c}{OOD} & \multicolumn{2}{c}{Recorded} \\
\cmidrule(lr){2-4}\cmidrule(lr){5-7}\cmidrule(lr){8-10}\cmidrule(l){11-12}
\textbf{STA} & FAD$\downarrow$ & CLAP$\uparrow$ & OVL$\uparrow$ & FAD$\downarrow$ & CLAP$\uparrow$ & OVL$\uparrow$ & FAD$\downarrow$ & CLAP$\uparrow$ & OVL$\uparrow$ & FAD$\downarrow$ & CLAP$\uparrow$ \\
\midrule
STAR~\cite{xie2025starspeechtoaudiogenerationrepresentation} & 2.23 & \textbf{0.422} & \underline{2.68} & \underline{2.23} & \underline{0.382} & 2.36 & 2.16 & 0.386 & 2.54 & 1.98 & 0.394 \\
ST5 + FM & \textbf{1.72} & 0.412 & 2.54 & 2.28 & 0.339 & 1.92 & \textbf{1.76} & 0.387 & 2.32 & 2.23 & 0.367 \\
SE $\rightarrow$ ST5 + FM & \underline{1.92} & 0.405 & 2.51 & 2.35 & 0.367 & \underline{2.90} & \underline{1.79} & \underline{0.403} & \textbf{2.72} & \underline{1.85} & \underline{0.402} \\
AlignAudio-STA & 1.96 & \underline{0.418} & \textbf{2.86} & \textbf{1.88} & \textbf{0.421} & \textbf{3.12} & 1.84 & \textbf{0.421} & \underline{2.62} & \textbf{1.78} & \textbf{0.414} \\
\cmidrule(lr){1-12}
EzAudio~\cite{hai2024ezaudio} (GT text) & 1.68 & 0.427 & 3.02 & -- & -- & -- & -- & -- & -- & -- & -- \\
\midrule
\textbf{VC} & WER$\downarrow$ & SECS$\uparrow$ & NMOS$\uparrow$ & WER$\downarrow$ & SECS$\uparrow$ & NMOS$\uparrow$ & WER$\downarrow$ & SECS$\uparrow$ & NMOS$\uparrow$ & WER$\downarrow$ & SECS$\uparrow$ \\
\midrule
FreeVC~\cite{li2023freevc} & 5.21 & 54.62 & 4.20 & 36.36 & 34.45 & 3.26 & 33.74 & 36.02 & 3.34 & 37.84 & 35.46 \\
SE $\rightarrow$ FreeVC & 5.81 & 52.88 & 4.12 & \underline{31.84} & 40.71 & \underline{3.48} & 29.62 & 42.15 & \textbf{3.54} & 28.49 & 43.25 \\
DiffVC~\cite{popov2022diffusionvc} & 8.27 & \textbf{62.49} & 4.08 & 84.57 & 42.61 & 2.11 & 78.95 & 44.83 & 2.56 & 57.42 & 42.49 \\
Noro~\cite{he2025noro} & 5.12 & 55.23 & 4.39 & 45.27 & \underline{50.52} & 3.04 & 42.18 & \underline{53.13} & 3.08 & 45.84 & \underline{54.13} \\
MeanVC~\cite{ma2025meanvc} & \underline{4.69} & 57.49 & \underline{4.42} & 108.93 & 28.77 & 2.08 & 101.47 & 30.52 & 2.14 & 78.50 & 31.59 \\
AlignAudio-VC* & 5.25 & \underline{58.12} & \textbf{4.55} & 32.36 & \textbf{53.44} & \textbf{3.51} & \underline{29.41} & \textbf{54.76} & \underline{3.36} & \underline{28.48} & \textbf{56.14} \\
AlignAudio-VC & \textbf{4.26} & 54.83 & \underline{4.43} & \textbf{27.56} & 49.82 & 3.07 & \textbf{24.93} & 50.91 & 3.13 & \textbf{25.76} & 53.46 \\
\bottomrule
\end{tabular*}
\end{table*}

\subsection{Experimental Setup}
\label{sec:experimental_setup}


\noindent\textbf{Datasets.}
For STA, we generate synthetic speech versions of AudioCaps~\cite{kim2019audiocaps} captions using the SpeechT5 text-to-speech model, following the official AudioCaps splits ($\sim$50k clips in total).
For VC, we use LibriTTS~\cite{Zen2019LibriTTSAC} for training ($\sim$585 hours) and evaluate on \textit{test-clean}, whose speakers are completely disjoint from the training set.
For real-world evaluation, we record speech from a speaker reading AudioCaps captions under five acoustic conditions: clean, room, street, cafe, and subway.
Ten utterances are randomly recorded for each condition.
The same recordings are used as speech conditions for STA and source speech for VC.

\noindent\textbf{Noise conditions.}
We evaluate robustness under both synthetic and real-world noise conditions.
MUSAN~\cite{snyder2015musan} and RIRS\_NOISES~\cite{ko2017study} are used to simulate music, environmental noise, and reverberation during training, where each augmentation type is applied with a probability of 0.3 and SNRs are randomly sampled from 5--20 dB.
For VC, augmentation is applied to the source speech, target reference speech, or both with probabilities of 0.3, 0.3, and 0.4, respectively.
For OOD evaluation, DEMAND~\cite{thiemann2013diverse} noises are added to the test sets with randomly sampled SNRs from -5--20 dB.
The real-world recordings are directly evaluated without additional noise mixing to reflect uncontrolled acoustic variations.

\noindent\textbf{Implementation.}
For STA, the generative core is a 24-layer DiT with 16 attention heads and a hidden dimension of 1024.
The weighting coefficient $\lambda$ is set to 0.9 to prioritize the stabilization of clean speech representations.
The alignment coefficient is $\alpha=0.05$, and the scheduling function $\omega(t)$ uses $k=20$ and $b=0.3$.
For VC, the core architecture adopts the framework of MeanVC~\cite{ma2025meanvc}.
Since only a few steps are used for sampling, timestep-dependent weighting is disabled ($\omega(t)\equiv1$).
We further adapt Flow-GRPO~\cite{liu2025flowgrpo} after the two-stage training and report results with and without GRPO.

\noindent\textbf{Baselines.}
We compare AlignAudio with representative open-source baselines for both STA and VC tasks.
For STA, we include speech-conditioned generation systems (STAR and ST5 + FM) and an enhancement-assisted pipeline (SE $\rightarrow$ ST5 + FM).
EzAudio with ground-truth text is additionally included as an oracle text-conditioned reference.
For VC, we compare with representative systems including FreeVC, DiffVC, Noro, and MeanVC, covering non-diffusion, diffusion-based, and noise-robust voice conversion approaches; an enhancement-assisted FreeVC pipeline (SE $\rightarrow$ FreeVC) is also considered.
All baselines are evaluated using their official implementations without retraining.

\noindent\textbf{Metrics.}
For STA, we report FAD and CLAP score as objective metrics, and Overall Audio Quality (OVL) as the subjective metric.
For VC, we report WER computed using Whisper~\cite{radford2023robust}, SECS using ECAPA-TDNN~\cite{desplanques20_interspeech} to evaluate timbre similarity, and Naturalness Mean Opinion Score (NMOS) as the subjective metric.
The subjective evaluation was carried out by 10 listeners for each task.

\vspace{-1em}
\subsection{Main Results}

\noindent\textbf{STA.}
Table~\ref{tab:main_results_combined} shows that ST5 + FM's CLAP drops from 0.412 to 0.339 under noise. Front-end SE partially recovers it to 0.367, whereas AlignAudio-STA reaches 0.421 and achieves the best noisy FAD and OVL among speech-conditioned systems. AlignAudio-STA's CLAP remains stable across clean (0.418), noisy (0.421), and OOD (0.421) conditions, supporting semantic robustness beyond the training noise sources. These improvements support explicit robustness training within the generation system to preserve both semantic fidelity and perceptual quality.

\noindent\textbf{VC.}
FreeVC and MeanVC show substantial degradation in both content preservation and speaker similarity under noisy and OOD conditions. Front-end SE improves both metrics for FreeVC, while Noro preserves speaker similarity well. AlignAudio-VC* achieves the highest SECS under both conditions, exceeding SE $\rightarrow$ FreeVC by 12.73 and 12.61 percentage points, respectively, with comparable WER. Compared with Noro, AlignAudio-VC* achieves lower WER under noisy (32.36\% vs.\ 45.27\%) and OOD (29.41\% vs.\ 42.18\%) conditions, while also attaining higher SECS. These results support dual alignment for jointly preserving linguistic content and target-speaker fidelity. Adding Flow-GRPO further yields the lowest WER across all four conditions, while AlignAudio-VC* retains higher speaker similarity and naturalness.

\noindent\textbf{Recorded speech.}
The recorded evaluation examines whether gains obtained under simulated corruption persist in real acoustic environments.
On the same human recordings, AlignAudio-STA leads in FAD and CLAP, while the full AlignAudio-VC achieves the lowest WER and AlignAudio-VC* the highest SECS.
Their advantage over the SE pipelines supports the usefulness of the framework under naturally recorded interference in both tasks.

\vspace{-1em}
\subsection{Ablation Studies}

\begin{table}[tb]
\centering
\setlength{\tabcolsep}{2pt}
\renewcommand{\arraystretch}{1.05}
\caption{Ablation under noisy conditions; VC uses AlignAudio-VC* (without GRPO). Stages 1/2: representation/generation alignment. Full STA uses Early-$t$; weighting variants apply only to STA. \textbf{Bold}/\textit{italics} mark the best/second-best per column.}
\label{tab:ablation_combined}
\begin{tabular*}{\columnwidth}{@{\extracolsep{\fill}}lcccc@{}}
\toprule
Configuration & \multicolumn{2}{c}{STA} & \multicolumn{2}{c}{VC} \\
\cmidrule(lr){2-3}\cmidrule(l){4-5}
& FAD$\downarrow$ & CLAP$\uparrow$ & WER$\downarrow$ & SECS$\uparrow$ \\
\midrule
Full dual-alignment & \textbf{1.88} & 0.421 & \textbf{32.36} & \textbf{53.44} \\
w/o Stage 1 & 2.24 & 0.384 & 76.55 & 30.57 \\
w/o Stage 2 & 2.28 & 0.387 & 87.54 & 32.89 \\
w/o both stages & 2.28 & 0.339 & 108.93 & 28.77 \\
\midrule
Noisy augmentation & 2.11 & 0.398 & 56.28 & 40.16 \\
Joint training & 2.07 & 0.403 & \textit{35.16} & \textit{46.29} \\
\midrule
Uniform & 1.98 & \textbf{0.424} & -- & -- \\
Mid-$t$ & 2.36 & \textit{0.422} & -- & -- \\
Late-$t$ & \textit{1.90} & 0.418 & -- & -- \\
\bottomrule
\end{tabular*}
\end{table}

Table~\ref{tab:ablation_combined} demonstrates the complementary roles of the two alignment stages. Adding generation-level alignment reduces VC WER from 87.54\% to 32.36\% and increases STA CLAP from 0.387 to 0.421. Removing representation-level alignment also degrades both tasks. Full dual alignment further outperforms noisy augmentation and joint training, supporting sequential adaptation of conditioning representations and generation dynamics.

Early-$t$ achieves the lowest FAD (1.88) among the evaluated weighting schedules, while CLAP varies within 0.418--0.424. These results support the early-weighted configuration for improving acoustic fidelity while preserving semantic alignment.

\vspace{-1em}
\section{Conclusion}
\vspace{-0.5em}
In this paper, we propose AlignAudio, a framework that improves the noise robustness of speech-conditioned generation by aligning conditioning representations and generation dynamics. The framework supports single- and multi-stream speech conditioning with flow matching and MeanFlow, and integrates Flow-GRPO for VC post-training. Experiments on STA and VC demonstrate improved conditioning fidelity under synthetic noise and real-world recorded interference.
Future work will incorporate larger and more diverse real-world noisy datasets, including spontaneous speech, and extend our framework to a wider array of downstream audio applications.

\vfill\pagebreak
\section{Compliance with Ethical Standards}
This study involved subjective evaluation of audio samples by human listeners. No personally identifiable information was collected, and the evaluation was conducted anonymously. Ethical approval was not required according to the policy of the authors' institution.
\bibliographystyle{IEEEbib}
\bibliography{strings,refs}

\end{document}